\documentclass[twocolumn,prl,aps]{revtex4}
\usepackage{floatflt}
\usepackage{psfrag}
\usepackage{amsmath}   
\usepackage{amssymb}
\usepackage{amsfonts}
\usepackage{mathrsfs}
\usepackage{graphicx}

\begin{document}
\title
{Novel 2D Plasmon Induced by Metal Proximity}

\author{V.~M.~Muravev, P.~A.~Gusikhin, A.~M.~Zarezin, I.~V.~Andreev, S.~I.~Gubarev, I.~V.~Kukushkin}
\affiliation{Institute of Solid State Physics, RAS, Chernogolovka, 142432 Russia}
\date{\today}

\date{\today}

\begin{abstract}
A new electromagnetic plasma mode has been discovered in the hybrid system formed by a highly conductive gate strip placed in proximity to the two-dimensional electron system. The new plasmon mode propagates along the gate strip with no potential nodes present in transverse direction. Its unique spectrum combines characteristic features of both gated and ungated 2D plasmons. The new plasma excitation has been found to exhibit anomalously strong interaction with light.
\end{abstract}

\maketitle

The problem of electromagnetic waves propagating along metallic wires was treated rigorously by Sommerfeld more than 100 years ago~\cite{Sommerfeld}. He demonstrated that plasmon polariton waves travel along the wires at the speed of light.  In fact, it is these plasmon polariton waves that carry alternating signal along the modern power lines. Recently, it has been  shown theoretically that if a highly conducting wire is placed very close to a much less conducting two-dimensional electron system (2DES), the hybrid system may support a new electromagnetic plasma mode~\cite{Volkov:19}. Remarkably, this proximity plasmon mode differs significantly from both gated and ungated 2D plasma modes that have been known in the literature for over 50 years~\cite{Stern, Chaplik:72, Grimes:76, Allen:77, Theis:77}. Its dispersion is described by the following expression:
\begin{equation}
\omega_{\rm pr}(q) = \sqrt{\frac{2 n_s e^2 h}{m^{\ast} \varepsilon \varepsilon_0} \frac{q}{W}}  \qquad (qW \ll 1),
\label{newplasmon}
\end{equation} 
where $n_s$ is the 2D electron density, $m^{\ast}$ is the effective electron mass, $h$ is the distance between the gate and 2DES, $W$ is the width of the gate strip, $q$ is the plasmon wave vector directed along the infinite gate and $\varepsilon$ is the effective dielectric permittivity of the surrounding medium. Remarkably, the spectrum  (\ref{newplasmon}) combines fundamental features of both gated ($\omega_g \propto \sqrt{h}$) and ungated ($\omega_p \propto \sqrt{q}$) plasmons.

In the presence of external magnetic field, $B$, normal to the plane of the 2DES, there occurs hybridization between the cyclotron and the plasma motions, which leads to the proximity plasmon magnetodispersion:
\begin{equation}
\omega = \sqrt{\omega_{\rm pr}^2 + \omega_c^2},
\label{Bplasmon}
\end{equation} 
where $\omega_c = eB/m^{\ast}$ is the electron cyclotron frequency. It has been shown in~\cite{Allen:83, Volkov:EMP} that in a finite 2DES, two different types of plasmon modes exist --- the cyclotron magnetoplasmon and the one-way edge magnetoplasmon. Importantly, the new proximity plasmon excitation does not exhibit the edge magnetoplasmon mode in the presence of magnetic field. For quantitative treatment of the novel 2D plasmon mode the reader is referred to Supplemental Material~I.  

There are two reasons for which the observation of proximity plasma mode has been hindered over the past $50$ years. Foremost, it is due to the fact that proximity plasmon has no charge density nodes in the direction perpendicular to the metal strip, which naturally makes the mode dark to the electromagnetic wave with transversely polarized electric field --- a typical configuration used in all pioneering experiments in the field of 2D plasmonics~\cite{Allen:77, Theis:77, Heitmann:84, Heitmann:91}. Secondly, the most common geometry considered in theoretical analysis has been that of a finite 2DES with an infinite 2D screening gate~\cite{Fetter:86}. The configuration which is, in a sense, reciprocal to the one necessary to observe the proximity plasmon. In presented experiments, we succeeded in making manifest the long lost plasmon excitation. 

The experiments were conducted on a single $30$~nm wide GaAs/AlGaAs quantum well structure with electron density in the range of $n_s=(2.2 - 2.8)\times10^{11}~\text{cm}^{-2}$. The quantum well was located at a distance of $h=4400$~\AA~below the crystal surface. Based on the transport measurements, the electron mobility at $T=1.5$~K was estimated to be $\mu=5\times10^6~\text{cm}^2/\text{V$\cdot$s}$. To enable excitation of plasma waves, a metallic gate was lithographically formed on the top surface of the sample, as illustrated by the inset in Fig.~1. The width and the length of the gate were varied from $W=20$~$\mu$m to $100$~$\mu$m and from $L=0.5$~mm to $1.7$~mm, respectively. Grounding contacts were fabricated on both sides of the gate, $0.2 - 0.4$~mm away from the strip edge. The microwave radiation was guided into the cryostat through a coaxial cable and then coupled to the gate strip by means of a coplanar waveguide transmission line. The microwave frequency was varied in the range of $1$ to $40$~GHz. In order to detect microwave absorption, we employed a non-invasive optical technique~\cite{Ashkinadze, Kukushkin:02}. The technique is based on high sensitivity of recombinant photoluminescence spectrum of 2D electrons to the electron temperature. The photoluminescence spectrum was recorded with and without the excitation microwave radiation and then the absolute value of the difference between the two spectra was integrated over the entire spectral range. Since the resultant integral is directly proportional to the change in the 2D electron temperature, it was used as a measure of microwave absorption. Sample was immersed in a liquid helium cryostat with superconducting coil. The superconducting coil was used to produce magnetic field ($B=0$\,--\,$2$~T) normal to the sample surface. All the experiments were performed at a temperature of $T=1.5$~K.

\begin{figure}[!t]
\includegraphics[width=\linewidth]{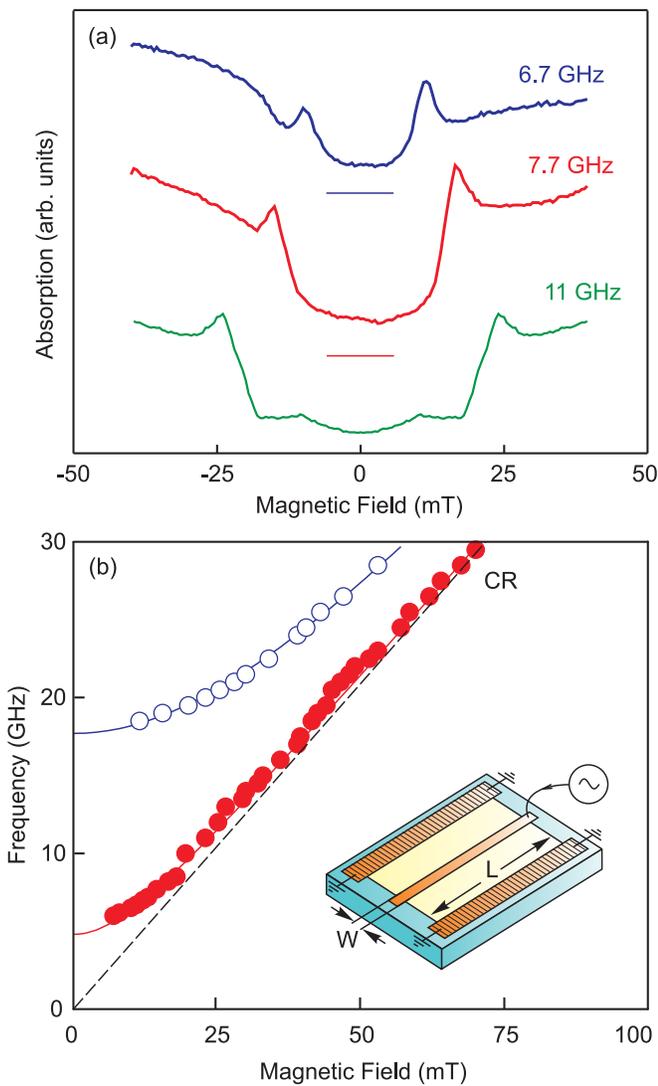}
\caption{(a) Dependencies of microwave absorption on magnetic field measured at frequencies of $6.7$, $7.7$ and $11$~GHz. The curves are normalized and offset for the purpose of clarity. Zero signal levels are indicated by short horizontal lines. (b) Magnetodispersion of the new proximity plasmon mode with transverse wave numbers $N=0$ (red circles) and $N=2$ (empty circles). The theoretical curves calculated for the two modes are plotted with solid red and blue lines, respectively. The inset displays a schematic view of the device with excitation geometry.} 
\label{fig1}
\end{figure} 

Figure~1~(a) illustrates microwave absorption measured in 2DES at $6.7$, $7.7$ and $11.0$~GHz as a function of magnetic field $B$. For these measurements there was used a structure with the gate strip dimensions $L=0.5$~mm and $W=0.1$~mm, and the 2D electron density, $n_s=2.7 \times 10^{11}~\text{cm}^{-2}$. The plotted data clearly indicate that there exists a pronounced absorption peak at each excitation frequency. The peak exhibits shift to higher values of magnetic field with microwave frequency increase. It was unexpected to find that the resonance arises at approximately $5$~GHz, which is considerably lower than predicted frequency of a gated plasmon~\cite{Chaplik:72}:

\begin{equation}
\omega_g(q) = \sqrt{\frac{n_s e^2 h}{m^{\ast} \varepsilon \varepsilon_0} } \, q  \qquad (qh \ll 1).
\label{scrplasmon}
\end{equation} 

According to Eq.~(3), the frequency of the lowest gated plasmon mode with wave vector~$q=\pi N/ W$ ($N = 1$) is equal to $10.4$~GHz. Therefore, experimental results presented in Fig.~1(a) demonstrate that we have, in fact, observed a new 2D plasma excitation induced in 2DES by metal proximity. 

Figure~1(b) displays the measured magnetodispersion of the new plasma mode. The red circles in the figure represent the data for the fundamental longitudinal proximity mode with transverse wave number $N=0$. This set of data is closely approximated by the quadratic function from~Eq.~(\ref{Bplasmon}) plotted with red line. In the strong magnetic-field limit, the data tends to the cyclotron frequency asymptote shown with dashed line. From the fitted curve, the mode can be extrapolated to yield the plasma frequency at $B=0$, $f_p(0)=4.8$~GHz. This value is in excellent agreement with the theoretical prediction obtained from Eq.~(\ref{newplasmon}) for $\varepsilon = 7.8$. This value of effective dielectric permittivity is very close to the average of the permittivity of GaAs and that of free space, $\varepsilon \approx (1 + \varepsilon_{\rm GaAs})/2$ for $\varepsilon_{\rm GaAs} = 12.8$ and $\varepsilon_{\rm vacuum} = 1$.


Unlike the fundamental mode, the transverse plasma oscillations have $N \ge 1$ potential nodes across the strip. The spectrum of the transverse modes in the limit of $qW \ll 1$ has been found in~\cite{Volkov:19} as:
\begin{equation}
\omega^2= \omega_g (q_{\rm tr})^2 + \omega_{\rm pr} (q)^2 = \frac{n_s e^2 h}{m^{\ast} \varepsilon \varepsilon_0} \left( q_{\rm tr}^2 +  \frac{4}{W} q  \right),
\label{newpl_modes}
\end{equation} 
where $q_{\rm tr} = N \pi/W$ ($N=1, 2 \ldots$) is the transverse component of the wave vector. Furthermore, in the  long wavelength limit of $qW \ll 1$, the expression in~Eq.~(\ref{newpl_modes}) reduces to an ordinary form describing a gated plasmon mode Eq.~(\ref{scrplasmon}). It is this mode that has been observed in numerous experiments~\cite{Eisenstein:00, Muravev:07, Andress:12, Koppens:18}, whereas the very fundamental plasma excitation with $N=0$ has been overlooked.

In addition to the fundamental mode, Fig.~1(b) includes measured dispersion data for the transverse proximity plasma mode with $N = 2$ denoted by empty circles. Importantly, due to symmetric geometry of the E-field within the coplanar waveguide, used as an excitation feed, only the modes with even wave number, $N=2, 4 \ldots$, can be excited in the given setup. As shown in the figure, the data can be extrapolated to estimate the mode frequency at $B=0$ to be $18$~GHz. The theoretical prediction for the transverse wave number, $N=2$, based on~Eq.~(\ref{newpl_modes}) is found to be $21$~GHz. Such a minor discrepancy between experiment and the theoretical prediction can likely be ascribed to inaccurate description of the 2DES dielectric environment.

The most significant and remarkable feature of the newly discovered proximity plasmon mode is its square-root dispersion~\cite{Volkov:19}. This is counter-intuitive, considering that a mode having 1D nature emerges in a gated 2DES system. It would be expected that both of these factors should favor linear dispersion law~\cite{Kukushkin:05, Muravev:07}. Nevertheless, experimental results in Fig.~2 clearly indicate square root dispersion for the proximity plasmon mode. Therefore, in order to test the theory, the actual dispersion was measured for three samples of different strip length, $L=0.5, 1.0$ and $1.7$~mm and fixed gate width, $W=100$~$\mu$m. Fig.~2 displays the resultant experimental data, designated by red circles, along with the curve calculated using~Eq.~(\ref{newplasmon}), plotted with a solid red line. It is evident from the figure that experimental results confirm the theoretical prediction of the square-root dispersion. The inset to Fig.~2 shows the measured proximity plasmon frequency as a function of parameter $1/W$, in which case the measurements were performed on three samples of different gate width, $W=100, 50$ and $20$~$\mu$m and fixed gate length, $L=0.5$~mm. According to~Eq.~(\ref{newplasmon}), $\omega_{\rm pr}$ is linearly proportional to $1/\sqrt{W}$. Hence, the obtained data are clearly in close agreement with the theoretical curve plotted with solid blue line.

\begin{figure}[!t]
\includegraphics[width=\linewidth]{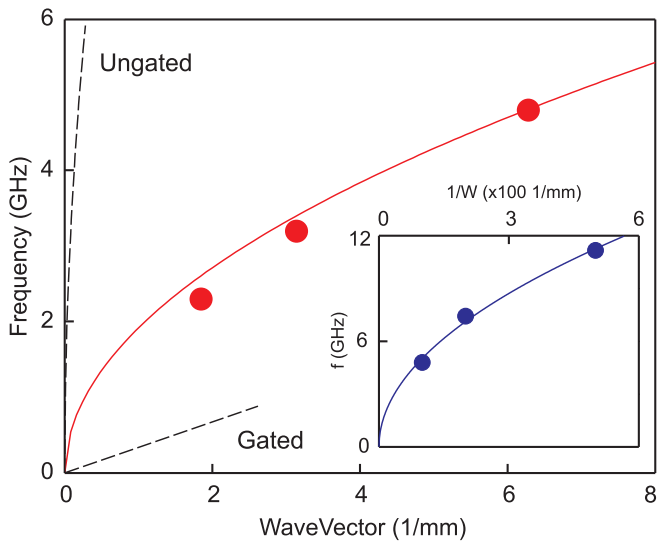}
\caption{Proximity plasmon dispersion measured for the gate strip width, $W=100$~$\mu$m, gate strip length, $L=0.5, 1.0$ and $1.7$~mm and electron density, $n_s=2.7\times10^{11}~\text{cm}^{-2}$. The theoretical curve is plotted with solid red line. For comparison, the calculated dispersion of ordinary gated and ungated 2D plasmons is indicated by dashed lines. The inset shows the dependence of plasmon frequency on $1/W$ with blue circles and solid line denoting measured and calculated data, respectively.} 
\label{fig2}
\end{figure}  

Concerning the experimental results in Fig.~2, it is worth noting that for the greatest gate strip length, $L=1.7$~mm, the measured data shows slight deviation from the square-root dispersion of approximately $\Delta \omega/\omega_{\rm pr} = 10$~\% . Such a reduction in plasmon resonant frequency for small wave vector values is a clear indication of hybridization between the plasma waves and the light. However, the magnitude of the observed hybridization far exceeds the level expected of the ordinary 2D plasmons~\cite{Kukushkin:03, Kukushkin:06, Muravev:18}. According to theory for the ungated 2D plasmon, $\Delta \omega/\omega_p \approx A^2/4$, where $A$ is a dimensionless retardation parameter defined as ratio of the plasmon frequency to that of light, given the same wave vector $q=\pi/L$~\cite{Muravev:18}. In essence, this parameter indicates the extent of the retardation effect. For the proximity plasmon mode under consideration, with $L=1.7$~mm, $W=100$~$\mu$m, $n_s=2.85 \times 10^{11}~\text{cm}^{-2}$, $A = 0.1$ which yields $\Delta \omega/\omega_{\rm pr} \approx 2.5 \times 10^{-3}$. Remarkably, the degree of coupling determined for the proximity plasmon experimentally is, in fact, $40$ times greater than that for the ungated 2D plasmon. 

One of the most attractive properties of 2D plasmons is their tunability. In our experiments, we were able to tune the 2D electron density in the given structure using photodepletion method~\cite{Kukushkin:89}. In Fig.~3 there are presented experimental results obtained for the sample with gate dimensions of $W=20$~$\mu$m and $L=0.5$~mm. The spectral data in Fig.~3(a) illustrates that with decrease in electron concentration from $n_s=2.3 \times 10^{11}~\text{cm}^{-2}$ to $0.6 \times 10^{11}~\text{cm}^{-2}$, the resonance peak shifts to the lower frequency. A more detailed dependence of the plasmon frequency on carrier density is shown in Fig.~3(b), where the measured data is compared against the theoretical curve that describes square-root dependence according to~Eq.~(\ref{newplasmon}). The measured and calculated data,  plotted with circles and solid line, respectively, show excellent agreement between experimental results and theory, demonstrating the possibility of tuning the speed of the new plasma mode over wide range.

\begin{figure}[!t]
\includegraphics[width=\linewidth]{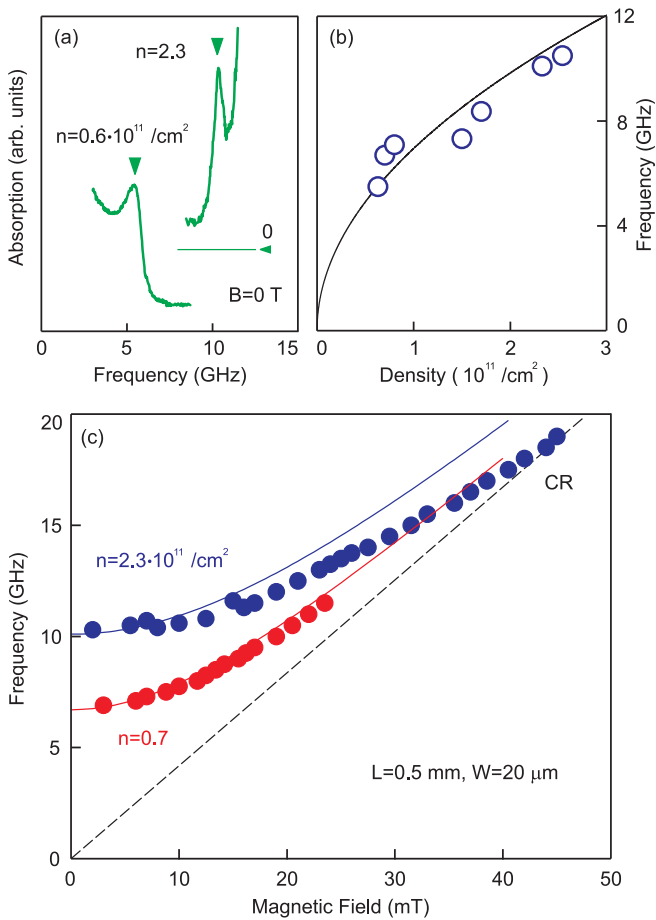}
\caption{(a) Microwave absorption spectra measured at $B=0$~T for electron concentrations of $n_s=0.6 \times 10^{11}~\text{cm}^{-2}$ and $2.3 \times 10^{11}~\text{cm}^{-2}$. The traces are offset for clarity. (b) Proximity plasmon frequency versus 2D electron density. An overlay plot of the measured data (empty circles) and the calculated curve (solid line). (c) Magnetodispersion for electron densities $n_s=0.7 \times 10^{11}~\text{cm}^{-2}$ and $2.3 \times 10^{11}~\text{cm}^{-2}$ with theoretical curves denoted by solid red and blue lines.} 
\label{fig3}
\end{figure} 

Figure~3(c) displays dependence of the proximity plasma mode frequency on magnetic field. The measurements were taken for the electron density of $n_s=0.7 \times 10^{11}~\text{cm}^{-2}$ (red circles) and $2.3 \times 10^{11}~\text{cm}^{-2}$ (blue circles). Experimental data closely follow the theoretical Pythagoras-like law, $\omega^2= \omega_p^2 + \omega_c^2$ (solid lines). We observe that in the high-density limit of $n_s=2.3 \times 10^{11}~\text{cm}^{-2}$ the magnetoplasma  mode intersects  the cyclotron resonance line. This again suggests the importance of the retardation effects viz. strong plasmon-photon coupling. Note that there was observed no influence of retardation effects on the proximity plasmon spectrum obtained for the sample with the same gate strip length, $L=0.5$~mm, but $5$ times greater width, $W=0.1$~mm, as shown in Fig.~1(b). Therefore, the gate width appears to be an additional factor controlling retardation for the novel proximity plasmon excitations. Found behavior differs dramatically from the coupling of light to the gated and ungated 2D plasmons. In fact, for gated 2D plasmons retardation effects are significantly suppressed and usually not observed at all~\cite{Chaplik:15}, whereas for ordinary ungated 2D plasmons retardation is dictated by only two parameters: the electron density and the sample size~\cite{Kukushkin:03, Kukushkin:06, Muravev:18}.

Notably, the overall structural geometry under consideration bears very close resemblance to that of a high-electron mobility transistor (HEMT). It has been shown that plasma oscillations in HEMT structures can be used for detection and generation of the terahertz radiation~\cite{Dyakonov:96, Shur:03, Knap:09}, which is based on the idea of compressing the incident radiation into highly-confined two-dimensional plasmons propagating in the transistor channel and then rectifying the induced \textit{ac} potential within the same device. However, despite the decades-long experimental efforts, terahertz plasmonic components are still far from their practical realization. Hence, the discovery of such a unique plasma mode that can exist in the HEMT device geometry opens new avenues for modern research in the field of terahertz electronics. 


In summary, we have for the first time discovered and experimentally investigated a new plasma excitation originating in the hybrid system formed by a metallic gate placed in close proximity to the 2DES. Extraordinarily, the measured spectrum of the new proximity plasmon excitation exhibits features of both gated ($\omega_g \propto \sqrt{h}$) and ungated ($\omega_p \propto \sqrt{q}$) 2D plasmons. Importantly, the observed plasmon mode has a wave vector along the gate strip and no potential nodes present in transverse direction. Therefore, the peculiar topology of the mode has rendered its detection impossible thus far because it cannot be excited with transversely polarized E-field - the common approach used in all pioneering experiments in 2D plasmonics. Furthermore, we have discovered that the new plasma excitation has anomalously strong interaction with light. This unique property makes the current discovery very promising for the development of sub-THz sensing instrumentation. In addition, being directly applicable to HEMT technology, it can lead to significant progress in developing practical plasmonic components in terahertz electronics. 

We thank V.A.~Volkov and A.A.~Zabolotnykh for the stimulating discussions. The authors gratefully acknowledge the financial support from the Russian Science Foundation (Grant No.~18-72-10072).

\end{document}


\title{Supplementary Material for\\ ``Novel 2D Plasmon Induced by Metal Proximity''}
\author{V.~M.~Muravev, P.~A.~Gusikhin, A.~M.~Zarezin, I.~V.~Andreev, S.~I.~Gubarev, I.~V.~Kukushkin}
\affiliation{Institute of Solid State Physics, RAS, Chernogolovka, 142432 Russia}

\date{\today}\maketitle

\section{\textrm{I}. Quantitative interpretation of the new 2D plasma mode}

To estimate the frequency of the new 2D plasma mode, the plasmon wave can be treated in terms of its equivalent $LC$ oscillator circuit~\cite{Aizin:12, Yoon:14}. Consider each plasmon wave fragment of length $\lambda_p=1/q$ as a separate oscillator. The kinetic energy of collective electron oscillations in such an oscillator can be modelled using the kinetic inductance of a non-magnetic origin, $L$. On the other hand, the electric potential energy associated with the Coulomb's restoring force driving local electrons into plasmonic oscillation can be modelled using the electrical capacitance, $C$. In the case under consideration, the capacitance of the plasmonic oscillator is determined by dimensions of the central metal strip (Fig.~1).

\begin{equation}
    C = \varepsilon_0 \varepsilon \, \frac{W \lambda_p}{d}.
\end{equation}

Then, the electrons capacitively accumulated under the central strip discharge through the regions adjacent to 2DES (Fig.~1). The kinetic energy, $E_k$, of the accelerating electrons is closely linked to the kinetic inductance. In fact, for the electron velocity, $v$, at a given time, the total kinetic energy, $K$, of the electrons in the 2DES can be expressed as $E_k = 2 \times m^{\ast}∗v^2/2 \times n_s \lambda_p^2$. 

Taking into account the fact that current flows on both sides of the strip up to the distance $\lambda_p$, the total current becomes $I=2 n_s e v \lambda_p$, which leads to $E_k=L I^2/2$, where $L$ is the total kinetic inductance of the plasmonic oscillator under consideration. As a result, we obtain: $L= L_k/2 = m^{\ast}/2 n_s e^2$.

Then, the resultant plasmon frequency can be defined as:
\begin{equation}
\omega_p=\frac{1}{\sqrt{L C}} = \sqrt{\frac{2 n_s e^2 d}{m^{\ast} \varepsilon \varepsilon_0} \frac{1}{W \lambda_p}}=
\sqrt{\frac{2 n_s e^2 d}{m^{\ast} \varepsilon \varepsilon_0 W} q}.
\end{equation}
Remarkably, our simple quantitative model leads to precise reproduction of the spectrum of the novel 2D plasmon Eq.~(1) calculated based on the exact theory~\cite{Volkov:19}.

\begin{figure}[b!]
\includegraphics[scale=0.8]{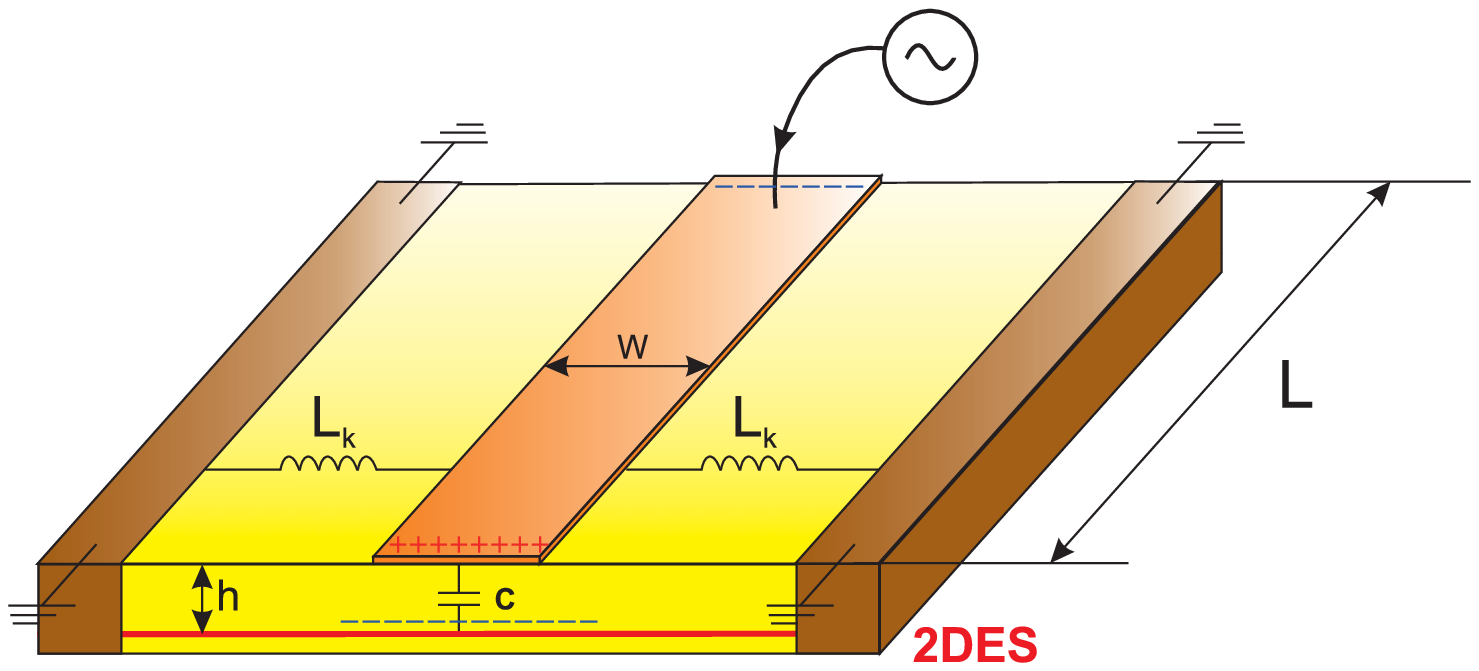}
\caption{Schematic drawing of the plasmon oscillator equivalent circuit.  Metallic gate of width $W$ and length $L$ is etched on the top surface of the crystal. Grounding contacts of the 2DES are added at each side of the gate strip. Quantum well is located at a distance $h$ below the sample surface. $C$ and $L_k$ are the effective capacitance  and the kinetic inductance of the plasmonic oscillator. The diagram is not drawn to scale.}
\label{image1}
\end{figure}